\documentclass{article}
\usepackage{spconf,amsmath,graphicx}
\usepackage{multirow}
\usepackage{amssymb,mathrsfs}
\usepackage{steinmetz}
\usepackage{mathabx}
\usepackage{mathtools}

\usepackage{enumitem}
\usepackage{url}
\usepackage{hyperref}

\title{R\MakeLowercase{e}SIFT: Reliability-Weighted SIFT-based Image Quality Assessment} 

\name{Dogancan Temel and Ghassan AlRegib}
\address{Center for Signal and Information Processing (CSIP)\\
School of Electrical and Computer Engineering\\
Georgia Institute of Technology, Atlanta, GA, 30332-0250 USA\\
\{cantemel,alregib\}@gatech.edu}
\begin{document}
%

\onecolumn 

\begin{description}[labelindent=1cm,leftmargin=3cm,style=multiline]

\item[\textbf{Citation}]{D. Temel and G. AlRegib, "ReSIFT: Reliability-weighted sift-based image quality assessment," 2016 IEEE International Conference on Image Processing (ICIP), Phoenix, AZ, 2016, pp. 2047-2051.
} \\

\item[\textbf{DOI}]{\url{https://doi.org/10.1109/ICIP.2016.7532718}} \\

\item[\textbf{Review}]{Date added to IEEE Xplore: 19 August 2016} \\

\item[\textbf{Code/Poster}]{\url{https://ghassanalregib.com/publications/}} \\

\item[\textbf{Bib}] {
@INPROCEEDINGS\{Temel2016\_ReSIFT,\\ 
author=\{D. Temel and G. AlRegib\},\\ 
booktitle=\{2016 IEEE International Conference on Image Processing (ICIP)\},\\ 
title=\{ReSIFT: Reliability-weighted sift-based image quality assessment\},\\ 
year=\{2016\},\\ 
pages=\{2047-2051\},\\
doi=\{10.1109/ICIP.2016.7532718\},\\ 
ISSN=\{2381-8549\}, \\
month=\{Sept\},\}\\
} \\

\item[\textbf{Copyright}]{\textcopyright 2016 IEEE. Personal use of this material is permitted. Permission from IEEE must be obtained for all other uses, in any current or future media, including reprinting/republishing this material for advertising or promotional purposes,
creating new collective works, for resale or redistribution to servers or lists, or reuse of any copyrighted component
of this work in other works. }

\item[\textbf{Contact}]{\href{mailto:alregib@gatech.edu}{alregib@gatech.edu}~~~~~~~\url{https://ghassanalregib.com/}\\ \href{mailto:dcantemel@gmail.com}{dcantemel@gmail.com}~~~~~~~\url{http://cantemel.com/}}
\end{description} 

\thispagestyle{empty}
\newpage
\clearpage

\twocolumn

\maketitle
\begin{abstract}
This paper presents a full-reference image quality estimator based on SIFT descriptor matching over reliability-weighted feature maps. Reliability assignment includes a smoothing operation, a transformation to perceptual color domain, a local normalization stage, and a spectral residual computation with global normalization. The proposed method \textbf{ReSIFT} is tested on the LIVE and the LIVE Multiply Distorted databases and compared  with 11 state-of-the-art full-reference quality estimators. In terms of the Pearson and the Spearman correlation, \textbf{ReSIFT} is the \emph{best performing} quality estimator in the overall databases. Moreover, \textbf{ReSIFT} is the best performing quality estimator in at least one distortion group in compression, noise, and blur category.

\end{abstract}
\section{Introduction}
\label{sec:intro}

Image qualiy assessment methods algorithmically evaluate quality of images and the definition of the quality depends on a target application. Mean squared error and peak signal-to-noise ratio (PSNR) are commonly used in image coding applications,  where the quality criterian is based on fidelity. Human visual system characteristics can be used to enhance fidelity metrics in terms of perceptual correlation as in PSNR-HVS \cite{Ponomarenko2007}, PSNR-HVS-M \cite{Ponomarenko2007}, PSNR-HA \cite{Ponomarenko2011}, and PSNR-HMA \cite{Ponomarenko2011}. 

Perception can also be introduced to image quality assessment algorithms using saliency-based approaches such as the spectral residual, which is the residual between a spectrum and an averaged spectrum. The spectral residual is based on the visual system characteristic that corresponds to suppressing responses to frequently occuring features and being sensitive to unexpected changes. The spectral residual approach is used to assign significance to gradient magnitudes to estimate image quality \cite{Zhang2012}. 

As an alternative to tracking changes in an intensity channel using a pixel-wise fidelity approach or focusing on sharp changes using a gradient magnitude operator, we can solely focus on discriminative features that are perceptually significant. The authors in \cite{Wen2014} use number of matched SIFT features to evaluate image quality. Instead of directly calculating the number of matched features over the whole image, the authors in \cite{Sun2014} compute number of SIFT features in a unit region on the first octave of the difference of Gaussian scale space of a preprocessed image to obtain a no-reference quality assessment. In \cite{Chen2013}, the authors  extract and match SIFT features to obtain affine transformation parameters and perform a reverse affine transformation to make full-reference quality assesment methods robust to deformations such as translation, rotation, scaling or skewing. In \cite{Decombas2012}, SIFT features are used to match objects in an original image to potentially deformed objects in a processed image. Standard deviation values between matched SIFT points are used to measure the non-rigid deformation level of an object and a structural-similarity index is computed between regions centered around SIFT features inside object boundaries. 

In the proposed work, we combine a smoothing operation, a color domain transformation, a normalization operation, and a spectral residual calculation to assign reliability to pixel maps. SIFT descriptors are extracted from the reliability-weighted pixel maps and percentile thresholds are computed from the distances between matched descriptors to obtain a quality indicator. Finally, a non-linear mapping is used to obtain the \texttt{ReSIFT} score. In Section \ref{sec:main}, a detailed description of the proposed quality estimator is provided including all the details to guarantee reproducibility in research. A validation analysis including the description of databases, compared quality estimators, and results is given in Section \ref{sec:val} and  we conclude our work in Section \ref{sec:conc}.

\begin{figure*}[t]
\centering
\includegraphics[width=0.95\linewidth, trim= 0mm 0mm 0mm 0mm]{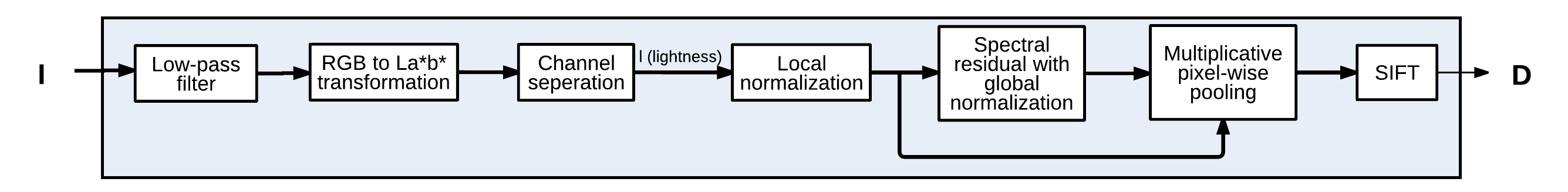}
\vspace{-5.0mm}
\caption{Reliability-weighted SIFT descriptor extraction}
\label{fig:BLeSS_Pipeline_part1}
\vspace{-5.0mm}
\end{figure*}

\section{Main}
\label{sec:main}

\subsection{Introduction to ReSIFT }
\label{subsec:main_intro}
\begin{figure}[htbp!]
\centering
\includegraphics[width=\linewidth, trim= 0mm 0mm 0mm 0mm]{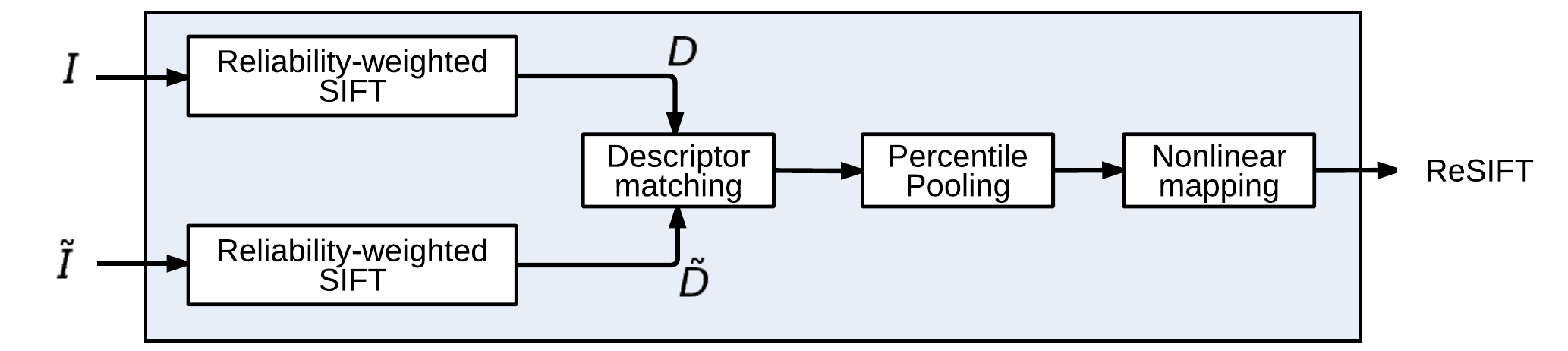}
\vspace{-8.0mm}
\caption{ReSIFT pipeline}
\label{fig:BLeSS_Pipeline_part2}
\vspace{-9.0mm}

\end{figure}

\vspace{-2.0mm}
The descriptor extraction process from an input image is summarized in Fig. \ref{fig:BLeSS_Pipeline_part1}. At first, an input image is smoothed out using a low-pass filter and then the smoothed image is transformed from the RGB to the La*b* color domain. A lightness map is separated and normalized locally, and a spectral residual with global normalization is computed from the locally normalized map. Then, the normalized lightness map and  the normalized spectral residual map are multiplicatively pooled pixel-wise and SIFT descriptors are extracted from the pooled map. The overall ReSIFT pipline is given in Fig. \ref{fig:BLeSS_Pipeline_part2}, where reliability-weighted SIFT descriptors are matched, pooled, and non-linearly mapped to obtain an estiamted quality score. The proposed quality estimator \texttt{ReSIFT} does not use any chroma information.

\vspace{-4.0mm}

\subsection{Low-pass filtering}
\label{subsec:lpf}
A rotationally symmetric Gaussian low-pass filter is used to smooth out an image. The cutoff frequency of the low pass filter is a function of the filter size ($f_{size}$) and the standard deviation ($f_{\sigma}$).
\vspace{-3.0mm}

\subsection{Perceptually uniform color space transformation}
\label{subsec:color}
\vspace{-2.0mm}
An RGB image is transformed into the perceptually uniform color space La*b* to separate lightness from chroma information. Transformation parameters include a transformation matrix $M$, CIE standard coefficients  $\kappa$ and $\epsilon$. A lightness channel is fed to the following blocks and chroma channels are not used due to lack of structural information.
\vspace{-3.0mm}

\subsection{Local normalization}
\label{subsec:local_norm}
\vspace{-2.0mm}
A local normalization operation over the lightness channel is performed by a mean subtraction and a divisive normalization operation. The local mean formulation is given as
\vspace{-2.0mm}
\begin{equation}
\label{eq:mean}
\mu[m,n]=\frac{1}{W^2}
\sum_{\hat{m} =floor(\frac{m}{W}) \cdot W+1}^{(floor(\frac{m}{W})+1) \cdot W} \sum_{\hat{n} =floor(\frac{n}{W}) \cdot W+1}^{(floor(\frac{n}{W})+1) \cdot W} l[\hat{m},\hat{n}], 
\end{equation}
where $m$ and $n$ are pixel indices, $l$ is the lightness channel, $W$ is the window size and floor is an operator that rounds a number to the next smaller integer. The standard deviation of each window is formualted as
\vspace{-3.0mm}
\begin{equation}
\label{eq:BSD2}
\begin{multlined}
\sigma[m,n]=\\  \sqrt{ \sum_{\hat{m} =floor(\frac{m}{W}) \cdot W+1}^{(floor(\frac{m}{W})+1) \cdot W} \sum_{\hat{n} =floor(\frac{n}{W}) \cdot W+1}^{(floor(\frac{n}{W})+1) \cdot W} \frac{(l[\hat{m},\hat{n}]- \mu[\hat{m},\hat{n}])^2}{W^2}}, 
\end{multlined}
\end{equation}
where  $\sigma$ is a standard deviation map. The local normalization operation is composed of two steps. First, a local mean ($\mu[m,n]$) is subtracted from each pixel in the lightness map. Then,  each mean shifted value is divided by a local standard deviation ($\sigma[m,n]$). 
\vspace{-3.0mm}

\subsection{Spectral residual with global normalization}
\label{subsec:SR}
\vspace{-2.0mm}
The normalized lightness map ($l_{norm}$) is transformed from the spatial domain to the frequency domain using the Fourier transform ($\mathscr{F}$). The magnitude component of the transformed map is expressed as 
\vspace{-3.0mm}
\begin{equation}
\vspace{-2.0mm}
\label{eq:SR1}
 \left|L\right|= \left| \left(\mathscr{F}  \left[ l_{norm}\right] \right)\right|, 
\end{equation}
where $\left| \cdot \right|$ is the magnitude operator and the phase component is given as 
\vspace{-3.0mm}
\begin{equation}
\label{eq:SR2}
\phase{L}= \phase{\left(\mathscr{F}  \left[ l_{norm}\right] \right)} , 
\end{equation}
where $\phase{\cdot}$ is the phase operator. The spectrum of a signal is computed by the log of the magnitude ($\log{ \left|L\right|}$) and the average spectrum is computed by convolving the spectrum with an averaging filter ($g$). The difference between the spectrum and the averaged spectrum results in the spectral residual, which is fomulated as 
\vspace{-5.0mm}
\begin{equation}
\label{eq:SR3}
SR(L)= \log{ \left|L\right|} - g \asterisk \log{ \left|L\right|} , 
\vspace{-2.0mm}
\end{equation}
where $\asterisk$ is the convolution operator and $g$ is the averaging filter. The spectral residual is combined with the phase of the normalized lightness map, and then an inverse Fourier transform is performed. In \cite{Hou2007}, a reconstructed image is used as a saliency map, which is referred as the unexpected portion of an image.

 A reconstruction operation using the spectral residual is formulated as
\vspace{-5.0mm}
\begin{equation}
\label{eq:SR4}
S= h \asterisk \mathscr{F}^{-1}\left[{SR(L)\phase{L}}  \right], 
\vspace{-2.0mm}
\end{equation}
where $\mathscr{F}^{-1}$ is the inverse Fourier transform operator, $h$ is the Gaussian low-pass filter used to smooth out a reconstructed map  and ${\left[ \cdot \phase{\cdot} \right]}$ is the representation of a signal in terms of its magnitude and its phase . The reconstructed map is globally normalized so that the pixel values are in between $0.0$ and $1.0$.
\vspace{-6.0mm}

\subsection{Multiplicative pixel-wise pooling}
\label{subsec:multip_pool}
\vspace{-2.0mm}
The locally normalized lightness map and the spectral residual-based resonstructed map are multiplicatively pooled pixel-wise to obtain a reliability-weighted lightness map.

\vspace{-2.0mm}
\subsection{Scale-invariant Feature Transform }
\label{subsec:main_sift}
\vspace{-2.0mm}
Scale-invariant feature transform (SIFT) is an algorithm \cite{Lowe04} that takes an image as an input and outputs
scale-invariant coordinates relative to local features. These features are desinged to be fully invariant to a translation, a scaling, and a rotation operation, and partially invariant to a change in the illumination and the camera viewpoint. The characteristics of these features resemble the properties of neurons in the inferrior temporal cortex of primate vision system \cite{Serre05}.

To obtain a SIFT feature vector, an input image is convolved using Gaussian kernels with nearby scales to obtain feature maps. A difference between these feature maps is calculated to obtain a difference of Gaussian (\texttt{DoG}) map. Each pixel in the \texttt{DoG} map is compared to its eight neighbors in the current map and nine neighbors in the scales above and below. If the center pixel is larger or smaller than all other pixels, it is selected as an extrema point. Rather than selecting a central point directly as a keypoint, a 3D quadratic function is fitted to sample points to obtain an interpolated location. Directions of local image gradients are used to assign orientations to the keypoints and a histogram orientation is formed using 36 bins that correspond to 360 degrees. Each sample in the historgram is weighted by a gradient magnitude and a Gaussian-weighted circular window. The highest peak and any other local peak within 80\% of the highest peak are used in the keypoint description. Therefore, keypoints with different orientations can have the same location and the same scale.

The location and the scale of a keypoint descriptor are used while computing the gradient magnitudes and the orientations. The magnitudes are weighted with a Gaussian function, whose standard deviation is a function of the descriptor window size. The orientation of a keypoint is used to rotate the computed gradient orientations relatively. A SIFT descriptor is obtained using a $4x4$ array of histograms with $8$ orientations, which lead to a feature vector of length $4x4x8=128$. In the proposed method \texttt{ReSIFT}, SIFT descriptors are extracted over the reliability-weighted lightness maps.
\vspace{-3.5mm}

\subsection{Descriptor matching}
\label{subsec:feature_match}
\vspace{-2.5mm}
The SIFT descriptors are matched based on the distance between them. Two descriptors are matched only if a threshold ($thresh$) times the Euclidean distance between them is less than the distance between that descriptor and other descriptors. Moreover, clusters of descriptors are analyzed based on their geometric characteristics to reject errenous matches. 
\vspace{-3.5mm}

\subsection{Percentile pooling}
\label{subsec:percentile_pooling}
\vspace{-2.0mm}
The SIFT descriptors are commonly used in object recognition, where target objects can be located anywhere in compared images. However, in case of image quality assessment, objects are only slighly rotated, translated or deformed because of distortions. Therefore, a mean distance between descriptors can be misleading. We use a percentile pooling strategy, which only requires a single parameter ($perc$), to obtain a threshold that only contains relativley small distances among matched descriptors. 
\vspace{-2.5mm}

\subsection{Nonlinear mapping}
\label{subsection:noninear_mapping}
\vspace{-2.0mm}
The percentile pooling threshold is proportional to the distance among the descriptors but the estimated quality is inversly proportional. Therefore, reciprocal of the percentile can be used to estiamte a quality score. An euclidean distance among $128D$ descriptors leads to values that are in the range of tens of thousands. To scale the range of the quality estimator, the percentile threshold is divided by a constant ($C_1$). We also add a constant ($C_2$) next to the divison in the reciprocal to avoid instabilities in case of small distance values. The nonlinear mapping function is given as
\vspace{-3.0mm}
\begin{equation}
\label{eq:nonlinear_mapping}
ReSIFT=\frac{1}{\frac{dist}{C_1}+C_2}
\vspace{-3.0mm}
\end{equation}
where $dist$ corresponds to the percentile threshold, $C_1$ and $C_2$ are coefficients and $ReSIFT$ is the estiamted quality score.
\vspace{-4.0mm}

\subsection{Parameter setup}
\vspace{-2.0mm}
All of the parameters used in the implementation of the proposed quality estimator are listed in Table \ref{tab_param}. VLFeat library \cite{vlfeat} is used for SIFT without any parameter tuning. The default values are used in the color space transformation and the spectral normalization. In the descriptor matching and the percentile pooling, the parameters are slightly tweaked. The low-pass filtering and the local normalization parameters are selected by visually assessing the feature maps and the nonlinear mapping values are set to fix the highest score to $100$ and to stabilize the proposed method.
\vspace{-4.0mm}

\begin{center}
\begin{table}[htbp!]
\scriptsize
\centering
\vspace{-4.0mm}
\caption{List of the parameters and their values in ReSIFT}
\label{tab_param}
\begin{tabular}{lcc}
\hline
\textbf{Section/Block}                      & \multicolumn{1}{l}{\textbf{Parameter}} & \multicolumn{1}{c}{\textbf{Value}}                                                \\ \hline
\multirow{2}{*}{2.2. Low-pass filtering} & $f_{size}$                              & 4                                                                                  \\ 
                                         & $f_{\sigma}$                            & 5                                                                                  \\ \hline
\multirow{5}{*}{2.3. Color space transformation} & \multirow{2}{*}{$\kappa$}               & \multirow{2}{*}{\begin{tabular}[c]{@{}c@{}}903.3\\ CIE standard\end{tabular}}      \\
                                         &                                         &                                                                                    \\  
                                         & $\epsilon$                              & 0.008856                                                                           \\  
                                         & \multirow{2}{*}{$M$}                    & \multirow{2}{*}{\begin{tabular}[c]{@{}c@{}}Adobe RGB  \\ stand. 1998\end{tabular}} \\
                                         &                                         &                                                                                    \\ \hline
2.4. Local normalization                 & $W$                                     & 20                                                                                 \\ \hline
\multirow{3}{*}{2.5. Spectral residual}  & $g_{size}$                              & 3                                                                                  \\  
                                         & $h_{size}$                              & 10                                                                                 \\  
                                         & $h_{\sigma}$                            & 3.8                                                                                \\ \hline
2.8. Descriptor matching                    & $thresh$                                & 1.4                                                                                \\ \hline
2.9. Percentile pooling                  & $perc$                                  & 5                                                                                  \\ \hline
\multirow{2}{*}{2.10. Nonlinear mapping} & $C_1$                                   & 100,000                                                                            \\ 
                                         & $C_2$                                   & 0.01                                                                               \\ \hline
\end{tabular}
\vspace{-6.0mm}
\end{table}
\end{center}

\vspace{-6.0mm}

\section{Validation}
\label{sec:val}

\vspace{-4.0mm}
\subsection{Databases}
\label{subsec:val_databases}
\vspace{-2.0mm}
\texttt{ReSIFT} is validated using the LIVE \cite{live2006} and the LIVE Multiply Distorted (MULTI) \cite{multi2012} databases. All of the distortion types in these databases can be grouped into four categories. \texttt{Compression} includes Jpeg, Jp2k, and Jpeg of blurred images.  \texttt{Noise} contains  white noise over reference images and white noise over blurred images. \texttt{Communication} includes Rayleigh fast-fading channel model errors and \texttt{Blur} consists of Gaussian blur. The number of images in each category is summarized in Table \ref{tab_db}. Blur-Noise and Blur-Jpeg are included in two different categories since pristine images are processed with both of the distortion types simultaneously.

\begin{table}[htbp!]
\scriptsize
\vspace{-2.50mm}
\centering
\caption{The number of distorted images with respect to degradation categories in each database}
\label{tab_db}
\begin{tabular}{cccc}
\hline
                    & {\bf LIVE} & {\bf MULTI}  & {\bf Total} \\ \hline
{\bf Compression}   & 460        & 225                  & 685        \\ 
{\bf Noise}         & 174        & 225                 & 399        \\ 
{\bf Communication} & 174        & -                    & 174         \\ 
{\bf Blur}          & 174        & 450                  & 624         \\ \hline
\end{tabular}
\end{table}
\vspace{-5.0mm}

\begin{center}

\begin{table*}[htbp!]
\scriptsize
\centering
\caption{Performance of IQA methods on different databases}
\label{tab_results_databases}
\begin{tabular}{ccccccccccccc}
\hline
               & \textbf{PSNR} & \textbf{PSNR-HA} & \textbf{PSNR-HMA} & \textbf{SSIM} & \textbf{MS-SSIM} & \textbf{CW-SSIM} & \textbf{IW-SSIM} & \textbf{SR-SIM} & \textbf{FSIM} & \textbf{FSIMc} & \textbf{PerSIM} & \textbf{ReSIFT}  \\ \hline
               & \multicolumn{12}{c}{\textbf{Pearson Correlation Coefficicent}}                                                                                                                                                                        \\ \hline
\textbf{LIVE}  & 0.927             &0.953                  &0.958                   &0.945               &0.946                  &0.872                  &0.951                  &0.945                 &0.949               &0.950                &0.955                 &\textbf{0.961}                                 \\ 
\textbf{MULTI} &0.739               &0.801                  &0.821                   &              0.812 & 0.802                 &0.379                  &0.847                  &0.888                 &0.818               & 0.821               & 0.852               &\textbf{0.906}                  \\ \hline

\textbf{}      & \multicolumn{12}{c}{\textbf{Spearman Correlation Coefficient}}                                                                                                                                                                        \\ \hline
\textbf{LIVE}  &0.909               &0.937                 &0.944                   &              0.949 & 0.951                 & 0.902                 & 0.960                 &0.955                 &  0.961          & 0.959              & 0.950                & \textbf{0.962} \\ 
\textbf{MULTI} &0.677               & 0.714                 & 0.742                  &              0.860  &0.836                  & 0.630                 &0.883                  &                0.866               & 0.863               & 0.866                &  0.818             & \textbf{0.887}              \\ \hline
\end{tabular}
\vspace{-4.0mm}
\end{table*}
\end{center}

\begin{center}
\begin{table*}[htbp!]
\scriptsize
\centering
\caption{Performance of best performing IQA methods on different distortion types}
\label{tab_results_types}
\begin{tabular}{c|c|cccc|ccccc}
\hline

\multirow{4}{*}{\bf Distortion Types} & \multirow{4}{*}{\bf Databases} & \multicolumn{4}{c|}{\multirow{2}{*}{\bf Pearson Correlation}}                                                            & \multicolumn{5}{c}{\multirow{2}{*}{\bf Spearman Correlation}}                                                            \\
                  &                   & \multicolumn{4}{l|}{}                                                                             & \multicolumn{5}{l}{}                                                                             \\ \cline{3-11} 
                  &                    & \multirow{2}{*}{\bf PSNR-HMA} & \multirow{2}{*}{\bf SR-SIM} & \multirow{2}{*}{\bf PerSIM}  & \multirow{2}{*}{\bf ReSIFT} & \multirow{2}{*}{\bf IW-SSIM} & \multirow{2}{*}{\bf FSIM} & \multirow{2}{*}{\bf FSIMc} & \multirow{2}{*}{\bf SR-SIM} & \multirow{2}{*}{\bf ReSIFT} \\
                  &                   &                                     &                   &                   &                   &                   &                   &                   &                   &                   \\ \hline

\multirow{3}{*}{{\bf Compression}}   & {\bf Jp2k {[}LIVE{]}}           &\textbf{0.982}&0.957&0.976&0.972&0.979&0.981&\textbf{0.982}&0.972&0.971          \\  
                                     & {\bf Jpeg {[}LIVE{]}}           &\textbf{0.972}&0.950&0.959&0.964&0.960&\textbf{0.962}&\textbf{0.962}&0.954&0.955     \\ 
                                     
                                     & {\bf Blur-Jpeg {[}MULTI{]}}     &0.838&0.904&0.864&\textbf{0.921}&0.869&0.854&0.855&0.862&\textbf{0.886} \\ \hline
\multirow{2}{*}{{\bf Noise}}         & {\bf Wn {[}LIVE{]}}             &0.985&0.974&0.968&\textbf{0.986}&0.982&0.980&0.979&0.983&\textbf{0.984}\\  
                                     & {\bf Blur-Noise {[}MULTI{]}}    &0.803&0.871&0.839&\textbf{0.897}&\textbf{0.893}&0.864&0.869&0.863&0.882\\ \hline
                                     
\multirow{1}{*}{{\bf Communication}} & {\bf FF {[}LIVE{]}}             &\textbf{0.954}&0.943&0.946&0.949&0.967&0.970&\textbf{0.971}&0.966&0.959\\ \hline
                                     
\multirow{3}{*}{{\bf Blur}}          & {\bf GBlur {[}LIVE{]}}          &0.950&0.949&0.967&\textbf{0.971}&\textbf{0.983}&\textbf{0.983}&\textbf{0.983}&0.978&0.979\\  
                                     
                                     & {\bf Blur-Jpeg {[}MULTI{]}}     &0.838&0.904&0.864&\textbf{0.921}&0.869&0.854&0.855&0.862&\textbf{0.886}\\ 
                                     & {\bf Blur-Noise {[}MULTI{]}}    &0.803&0.871&0.839&\textbf{0.897}&\textbf{0.893}&0.864&0.869&0.863&0.882\\ \hline

\end{tabular}
\end{table*}

\end{center}

\vspace{-20.0mm}

\subsection{Performance metrics}
\label{subsec:val_performance_metrics}
\vspace{-2.0mm}
The performance of the quality estimators are validated using the Pearson and the Spearman correlation coefficients. Since linearity-based Pearson correlation is sensitive to the range and to the distribution of the scores, a monotonic regression can be used before the correlation calculation. We use the monotonic formulation that can be expressed as 
\vspace{-3.0mm}
\begin{equation}
\label{eq:nonlinreg}
S=\beta_1 \left ( \frac{1}{1}-\frac{1}{2+exp(\beta_2(S_0 -\beta_3 ))} \right )+\beta_4 S_0 +\beta_5
\vspace{-2.0mm}
\end{equation}
where  $S_0$ is the input (raw value), $S$ is the regressed output, and $\beta$s are the tuning parameters that are set according to the relationship between the quality estimates and the mean opinion scores.

\subsection{State-of-the-art quality estimators}
\label{subsec:val_iqa}
In the performance comparison, we use full-reference quality estimators based on fidelity, perceptually-extended fidelity, structural similarity, feature similarity, and perceptual similarity, which include PSNR, PSNR-HA, PSNR-HMA, SSIM, MS-SSIM, CW-SSIM, IW-SSIM, SR-SIM, FSIM, FSIMc, and PerSIM. 

\vspace{-2.0mm}

\subsection{Results}
\label{subsec:val_results}
\begin{figure}[h!]

\begin{minipage}[b]{0.46\linewidth}

  \centering
\includegraphics[width=0.85\linewidth]{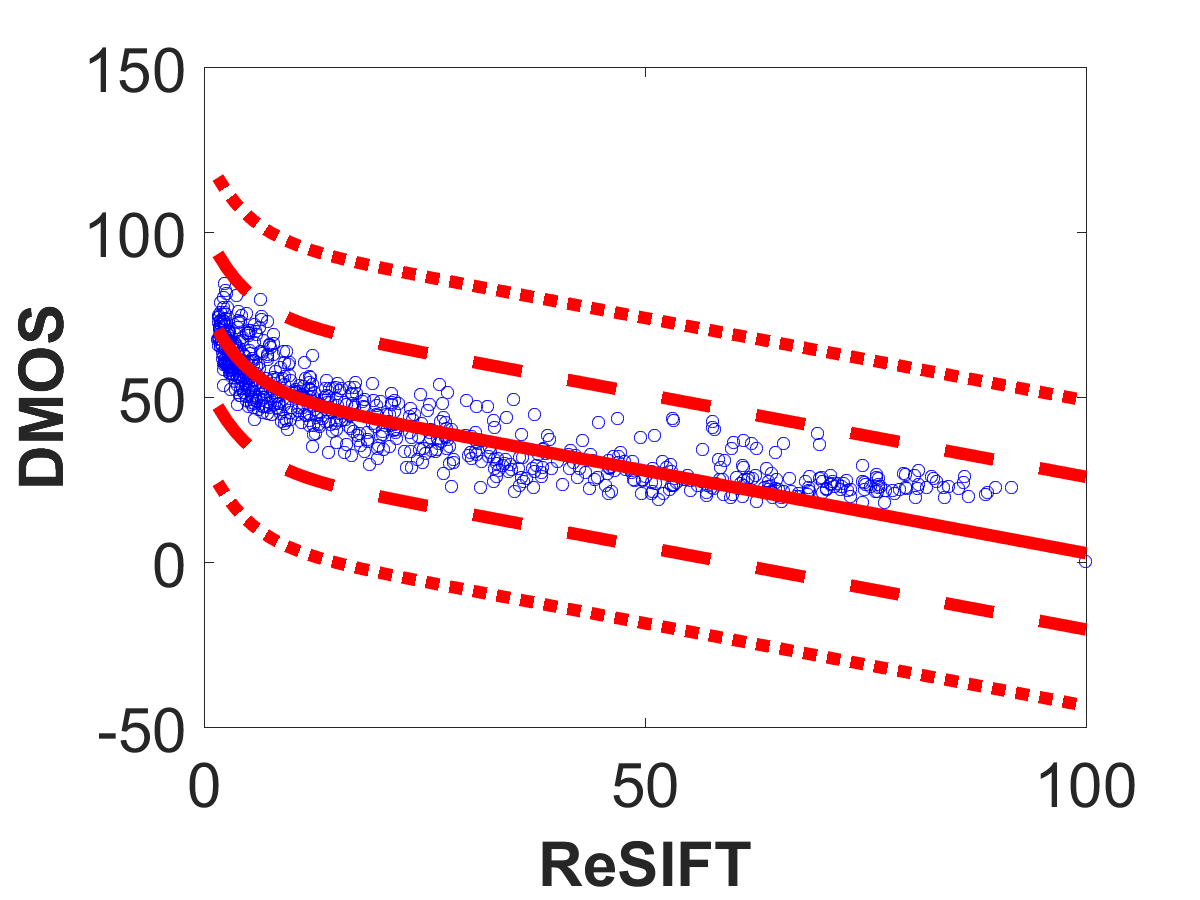}
  \centerline{\footnotesize{(a) LIVE-ReSIFT   } }
\end{minipage}
  \vspace{0.20cm}
\hfill
\begin{minipage}[b]{0.46\linewidth}
  \centering
\includegraphics[width=0.80\linewidth]{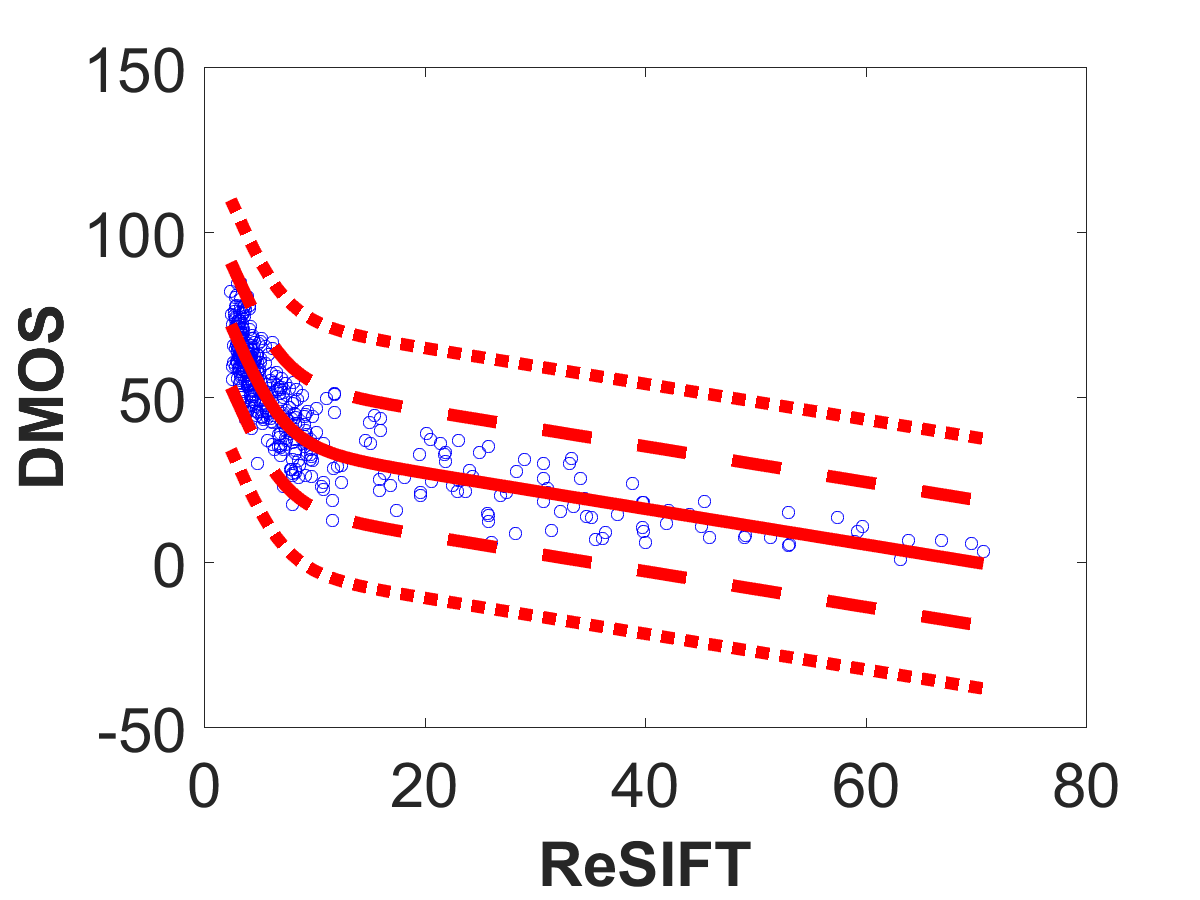}
  \centerline{\footnotesize{(b) MULTI-ReSIFT }}
\end{minipage}
\
\label{fig:Scatter}
\caption{Scatter plots of ReSIFT}
\vspace{-0.60 cm}
\end{figure}

\vspace{-2.0mm}
Performances of the quality estimators in overall databases are summarized in Table \ref{tab_results_databases}. \texttt{ReSIFT} is the highest performing quality estimator in the LIVE and the MULTI databases in terms of the Pearson and the Spearman correlation as highlighted with bold. The best performing estimators in the LIVE database are not significantly different from eachother but the difference becomes significant in the MULTI. Three best-performing quality estimators in each database and correalation categoy include PSNR-HMA, SR-SIM, PerSIM, IW-SSIM, FSIM, FSIMc, and \texttt{ReSIFT}. Performances of these quality estimators in various distortion types are given in Table \ref{tab_results_types}, where the highest performance in each category is highlighted. \texttt{ReSIFT} is the best performing quality estimator in at least one distortion group in compression, noise, and blur category. Scatter plots of \texttt{ReSIFT} corresponding to the LIVE and the MULTI databases are given in Fig. \ref{fig:Scatter}. The range of estimated quality socres are in between zero and hundred, and almost all the estimates are in the one standard deviation range with respect to the regression curve. Estimated quality scores use a wider quality range in the LIVE database compared to the MULTI database.

\section{Conclusion}
\label{sec:conc}
We proposed a reliability-weighted SIFT descriptor matching-based image quality estimator. The propsed method \texttt{ReSIFT} is used to quantify the perceptual quality of images under compression, noise, communication, and blur-based distortion types. When the LIVE and the LIVE Multiply Distorted image quality databases are considered, \texttt{ReSIFT} is the best performing quality estimator in at least one distortion group in compression, noise, and blur in terms of the Pearson and the Spearman correlation coefficients. Since proposed approach extracts  features all over an image and relies solely on a lightness channel, \texttt{ReSIFT} is inherently not designed for local, global, and color-based distortions. 

\newpage

\end{document}